\documentclass[12pt]{article}

\usepackage{amsxtra,amssymb,amsthm,amsmath,latexsym}
\usepackage{graphicx}
\usepackage{epsfig}
\usepackage{epstopdf}
\usepackage{booktabs}
\usepackage{multirow}
\usepackage{afterpage}
\usepackage{setspace}
\usepackage{microtype}
\usepackage{calc,fourier}
\usepackage[T1]{fontenc}
\usepackage[hidelinks]{hyperref}
\usepackage{float}
\usepackage[small]{caption}
\newtheorem{thm}{Theorem}[]

\numberwithin{equation}{section}

\newcommand{\bee}{\begin{equation*}}
\newcommand{\eee}{\end{equation*}}
\newcommand{\be}{\begin{equation}}
\newcommand{\ee}{\end{equation}}
\newcommand{\ba}{\begin{align}}
\newcommand{\ea}{\end{align}}

\newcommand{\RRR}{\mathbb{R}^3}

\title{Numerical Method for Solving Electromagnetic Scattering Problem by Many Small Impedance Bodies}
\author{N. T. Tran\footnote{Mailing address:  Mathematics Department, 138 Cardwell Hall, Manhattan, KS 66506} \\
\small Department of Mathematics\\[-0.8ex]
\small Kansas State University, Manhattan, KS 66506-2602, USA\\
\small \texttt{nhantran@math.ksu.edu}}

\date{}
\begin{document}
\maketitle

\begin{abstract}
In this paper, we study the problem of electromagnetic (EM) wave scattering by many small impedance bodies. A numerical method for solving this problem is presented. The problem is solved under the physical assumptions $a\ll d \ll \lambda$, where $a$ is the characteristic size of the bodies, $d$ is the minimal distance between neighboring bodies, $\lambda=2\pi/k$ is the wave length and $k$ is the wave number. This problem is solved asymptotically and numerical results for the cases of 125 and 1000 small bodies are presented to illustrate the method. Error estimates for the asymptotic solutions are also provided.
\end{abstract}



\section{Introduction} \label{sec0}

Electromagnetic scattering is the effect caused by EM waves such as light or radio waves hitting an object. The waves will then be scattered and the scattered field contains useful information about the object, see \cite{R278, Tsang2004, R470}. Electromagnetic scattering happens in many situations, for example, sun light scattered by atmosphere, radio waves scattered by buildings or planes, and so on.  The study of EM wave scattering is of great interest and importance since it helps advance many different fields ranging from Medical Technology to Computer Engineering, Geophysics, Photonics, and Military Technology, see \cite{Tsang2004, stavroulakis2013biological}. Unfortunately, most wave equations cannot be solved analytically to get a solution in a closed form. Therefore, numerical methods are sought to approximate the solution asymptotically. Unlike solving scalar wave scattering problem \cite{TMaterial, TFastScalar}, solving EM wave scattering is much more complicated and computationally expensive due to the vector nature of EM waves.

In \cite{R586} and \cite{RMaterial}, a theory for solving electromagnetic wave scattering problem by many small perfectly conducting and impedance bodies was developed. In \cite{TPerfect}, numerical methods for solving EM wave scattering by one and many small perfectly conducting are presented. In this paper, a numerical method for solving EM wave scattering by many small impedance bodies, based on the above theory, is described and tested. The problem is solved under the assumptions that the characteristic size $a$ of the bodies is much smaller than the distance $d$ between neighboring bodies, $d=O\left(a^\frac{2-\kappa}{3}\right)$ where $\kappa \in [0,1)$, and this distance $d$ is much less than the wave length $\lambda$, $ka \ll 1$ where $k$ is the wave number. The distribution of these small bodies is assumed to follow this law
\be 
    \mathcal{N}(\Delta)=\frac{1}{a^{2-\kappa}} \int_\Delta N(x)dx[1+o(1)], \quad a\to 0,
\ee
in which $\Delta$ is an arbitrary open subset of the domain $\Omega$ that contains all the small bodies, $\mathcal{N}(\Delta)$ is the number of the small bodies in $\Delta$,  and $N$ is the distribution function of the bodies
\be 
    N(x)\ge 0, \quad N(x) \in C(\Omega).
\ee
The boundary impedance of the bodies is of the form $\zeta = ha^{-\kappa}$, where $h$ is a continuous function such that Im$h\ge 0$, and $\kappa \in [0,1)$. The function $h$ and constant $\kappa$ can be chosen as desired.

To make the paper self-contained, the theory of EM wave scattering by one and many small impedance bodies is given in sections \ref{sec1} and \ref{sec2}. In section \ref{sec3}, the numerical method for solving the EM scattering problem is presented, the solutions of this problem are computed asymptotically and error analysis of the asymptotic solutions is also provided.

\section{Electromagnetic wave scattering by one small \\impedance body} \label{sec1}

Let $D$ be a bounded domain of one small body, $a$ be its radius, and $S$ be its smooth boundary, $S \in C^{1,\gamma}, \gamma \in (0,1]$. Assume that the dielectric permittivity $\epsilon$ and magnetic permittivity $\mu$ are constants. Let $E$ and $H$ denote the electric field and magnetic field, respectively. $E_0$ is the incident field and $v_E$ is the scattered field. The electromagnetic wave scattering by one small impedance body problem can be stated as follows
\begin{align}
            &\nabla \times E=i \omega \mu H, \quad \text{in }D':=\RRR \setminus D, \label{eq1.1.1} \\
            &\nabla \times H=-i \omega \epsilon E, \quad \text{in }D', \label{eq1.1.2} \\
            &[N,[E, N]]=\zeta [N, H], \quad \text{Re} \zeta \ge 0, \label{eq1.1.3} \\
            &E=E_0+v_E,  \label{eq1.1.4} \\
            &E_0=\mathcal{E} e^{ik\alpha\cdot x}, \quad \mathcal{E} \cdot \alpha=0,
            \quad \alpha \in S^2, \label{eq1.1.5} \\
            &\frac{\partial v_E}{\partial r}-ikv_E=o\left(\frac{1}{r}\right), \quad r:=|x|\to \infty, \label{eq1.1.6}
\end{align}
where $\omega>0$ is the frequency, $k=2\pi/\lambda=\omega\sqrt{\epsilon\mu}$ is the wave number, $ka \ll 1$, $\lambda$ is the wave length, $\zeta$ is the boundary impedance of the body, and $\alpha$ is a unit vector that indicates the direction of the incident wave $E_0$. This incident wave satisfies the relation $\nabla\cdot E_0=0$. The scattered field $v_E$ satisfies the radiation condition \eqref{eq1.1.6}. Here, $N$ is the outward pointing unit normal to the surface $S$.

It is known from \cite{RMaterial} that problem \eqref{eq1.1.1}-\eqref{eq1.1.6} has a unique solution and its solution is of the form
\be \label{eq1.3.2}
    E(x)=E_0(x)+\nabla \times \int_S g(x,t) J(t)dt, \quad g(x,t):=\frac{e^{ik|x-t|}}{4\pi |x-t|},
\ee
where $E_0$ is the incident plane wave defined in \eqref{eq1.1.5} and $J$ is an unknown pseudovector. $J$ is assumed to be tangential to $S$ and can be found from the impedance boundary condition \eqref{eq1.1.3}. Here $E$ is a vector in $\RRR$ and $\nabla \times E$ is a pseudovector.

Once we have $E$, $H$ can be found by the formula
\be \label{eq1.3.3}
    H=\frac{\nabla\times E}{i\omega \mu},
\ee

The asymptotic formula of $E$ when the radius $a$ of the body $D$ tends to zero is
\be \label{eq1.3.8}
    E(x)=E_0(x)+[\nabla_x g(x,x_1),Q],
\ee
where $|x-x_1|\gg a$, and the point $x_1$ is an arbitrary point inside the small body $D$, see \cite{RMaterial}. So, instead of finding $J$ to get $E$, we can just find one pseudovector $Q$
\be \label{eq1.3.7}
    Q=\int_S J(t)dt.
\ee

The analytical formula for $Q$ is derived in \cite{RMaterial} which can be summed up in the following theorem.
\begin{thm} \label{thm1.3.1}
    One has
    \be \label{eq1.3.15}
        Q=-\frac{\zeta |S|}{i\omega\mu}\tau \nabla \times E_0
    \ee
    where
    \be \label{eq1.3.16}
        \tau:=I_3-b, \quad b=(b_{jm}):=\frac{1}{|S|}\int_S N_j(s)N_m(s)ds,
    \ee
    and $|S|$ is the surface area of $S$.
\end{thm}
Here, $1 \le j, m \le 3$ correspond to $x,y$, and $z$ coordinates in $\RRR$, $I_3$ is a 3x3 identity matrix, and $N_j$, $1 \le j \le 3$, is the $j$-th component of the outer unit normal vector to the surface $S$.

\section{Electromagnetic wave scattering by many small \\impedance bodies} \label{sec2}

Now, consider a domain $\Omega$ containing $M$ small bodies $D_m$, $1\le m \le M$, and $S_m$ are their corresponding smooth boundaries. Let $D:=\bigcup_{m=1}^M D_m \subset \Omega$ and $D'$ be the complement of $D$ in $\RRR$. We assume that $S=\bigcup_{m=1}^M S_m \in C^{1,\gamma}, \gamma \in (0,1]$. We also assume that the dielectric permittivity $\epsilon$ and magnetic permittivity $\mu$ are constants. Let $E$ and $H$ denote the electric field and magnetic field, respectively. $E_0$ is the incident field and $v$ is the scattered field. The electromagnetic wave scattering by many small impedance bodies problem involves solving the following system
\begin{align}
            &\nabla \times E=i \omega \mu H, \quad \text{in }D':=\RRR \setminus D, \quad D:=\bigcup_{m=1}^M D_m, \label{eq2.1.1} \\
            &\nabla \times H=-i \omega \epsilon E, \quad \text{in }D', \label{eq2.1.2} \\
            &[N,[E, N]]=\zeta_m [N, H], \quad \text{on } S_m, \quad  1\le m\le M, \label{eq2.1.3} \\
            &E=E_0+v,  \label{eq2.1.4} \\
            &E_0=\mathcal{E} e^{ik\alpha\cdot x}, \quad \mathcal{E} \cdot \alpha=0,
            \quad \alpha \in S^2.
\end{align}
where $v$ satisfies the radiation condition \eqref{eq1.1.6}, $\omega>0$ is the frequency, $k=2\pi/\lambda$ is the wave number, $ka \ll 1$, $a:=\frac{1}{2}\max_m \text{diam}D_m$, $\alpha$ is a unit vector that indicates the direction of the incident wave $E_0$, and $\zeta_m$ is the boundary impedance of the body $D_m$. These $\zeta_m$'s are given by the following formula
\be \label{eq2.1.5}
    \zeta_m =\frac{h(x_m)}{a^\kappa}, \quad 0 \le \kappa <1, \quad x_m \in D_m,  \quad  1\le m\le M,
\ee
where $h(x)$ is a continuous function in a bounded domain $\Omega$,
\be \label{eq2.1.6}
    \text{Re} h(x) \ge 0, \quad \text{Im} \epsilon=\frac{\sigma}{\omega}\ge 0,
\ee
\be \label{eq2.1.7}
    \epsilon=\epsilon_0, \quad \mu=\mu_0 \quad \text{in } \Omega':=\RRR \setminus \Omega.
\ee
The distribution of small bodies $D_m$, $1 \le m \le M$, in $\Omega$ satisfies the following assumption
\be \label{eq2.1.8}
    \mathcal{N}(\Delta)=\frac{1}{a^{2-\kappa}} \int_\Delta N(x)dx[1+o(1)], \quad a\to 0,
\ee
where $\mathcal{N}(\Delta)$ is the number of small bodies in $\Delta$, $\Delta$ is an arbitrary open subset of $\Omega$,
\be \label{eq2.1.9}
    N(x)\ge 0, \quad N(x) \in C(\Omega)
\ee
and $\kappa \in [0,1)$ is the parameter from \eqref{eq2.1.5}.

Note that $E$ solves this equation
\be \label{eq2.1.11}
    \nabla\times\nabla\times E=k^2 E, \quad k^2=\omega^2\epsilon \mu,
\ee
if $\mu=$const.
Once we have $E$, then $H$ can be found from this relation
\be \label{eq2.1.10}
    H=\frac{\nabla\times E}{i\omega\mu}.
\ee
From \eqref{eq2.1.10} and \eqref{eq2.1.11}, one can get \eqref{eq2.1.2}. Thus, we need to find only $E$ which satisfies the boundary condition \eqref{eq2.1.3}. It was proved in \cite{RMaterial} that under the assumptions \eqref{eq2.1.6}, the problem \eqref{eq2.1.1}-\eqref{eq2.1.4} has a unique solution and its solution is of the form
\be \label{eq2.1.13}
    E(x)=E_0(x)+\sum_{m=1}^M \nabla\times \int_{S_m} g(x,t) J_m(t)dt.
\ee
 where
\be \label{eq2.1.15}
    Q_m:=\int_{S_m}J_m(t)dt.
\ee
When $a \to 0$, the asymptotic solution for the electric field is given by
\be \label{eq2.1.22}
    E(x)=E_0(x)+\sum_{m=1}^M [\nabla g(x,x_m), Q_m].
\ee
Therefore, instead of finding $J_m(t), 1 \le m \le M$, we can just find $Q_m$. The analytic formula for $Q_m$ is derived in \cite{RMaterial} by using formula \eqref{eq1.3.15} and replacing $E_0$ in this formula by the effective field $E_{em}$ acting on the m-th body
\be \label{eq2.1.24}
    Q_m=-\frac{\zeta_m|S_m|}{i\omega\mu}\tau_m \nabla\times E_{em}, \quad 1 \le m \le M.
\ee
The effective field acting on the m-th body is defined as
\be \label{eq2.1.23}
    E_e(x_m)=E_0(x_m)+\sum_{j\ne m}^M [\nabla g(x,x_j), Q_j]|_{x=x_m},
\ee
and $E_{em}:=E_{e}(x_m)$, $x_m$ is a point in $D_m$. When $a \to 0$, the effective field $E_e(x)$ is asymptotically equal to the field $E(x)$ in \eqref{eq2.1.22} as proved in \cite{RMaterial}.

From \eqref{eq2.1.5}, \eqref{eq2.1.24}, and \eqref{eq2.1.23}, one gets
\be \label{eq2.1.28}
    E_{em}=E_{0m}-\frac{ca^{2-\kappa}}{i\omega\mu}\sum_{j\ne m}^M [\nabla g(x,x_j)|_{x=x_m}, \tau \nabla \times E_{ej}]h_j,
\ee
where $c$ is a positive constant depending on the shape of the body $S_m$, $|S_m|=ca^2$, $\tau_m=\tau:=I_3-b$, and
\be
    b=(b_{jn}):=\frac{1}{|S_m|}\int_{S_m} N_j(s)N_n(s)ds.
\ee
Here, $1 \le j, n \le 3$ correspond to $x,y$, and $z$ coordinates in $\RRR$, $I_3$ is a 3x3 identity matrix, and $N_j$, $1 \le j \le 3$, is the $j$-th component of the outer unit normal vector to the surface $S_m$.

Our goal is to find $E_{em}$ in \eqref{eq2.1.28}. Take curl of \eqref{eq2.1.28}, set $x=x_j$, and let $A_m:=\nabla \times E_{em}$, we have
\be \label{eq2.1.29}
    A_j=A_{0j}-\frac{c a^{2-\kappa}}{i\omega\mu} \sum_{m\ne j}^M k^2 g(x_j,x_m)\tau A_m h_m + (\tau A_m \cdot \nabla_x)\nabla g(x,x_m)|_{x=x_j} h_m,
\ee
where $ 1 \le j \le M$, see \cite{RMaterial}. Solving this linear system gives us the curl of $E_{em}$, for $1 \le m \le M$.

The solution to the scattering problem \eqref{eq2.1.1}-\eqref{eq2.1.4} is described in the following theorem in \cite{RMaterial}.
\begin{thm} \label{thm2.1.2}
    The solution to the scattering problem \eqref{eq2.1.1}-\eqref{eq2.1.4} is given by formula \eqref{eq2.1.28} in which vectors $A_m:=\nabla \times E_{em}$ are found from linear algebraic system \eqref{eq2.1.29}.
\end{thm}

\section{Computing the solution of EM scattering problem by many small impedance bodies} \label{sec3}

In this section, we solve electromagnetic wave scattering by many small impedance bodies problem numerically under the theory described in section \ref{sec2}. In particular, we solve the following system for $A_j:=\nabla \times E_{ej}$, $1 \le j \le M$
\be \label{eq3.1.1}
    A_j=A_{0j}-\frac{c a^{2-\kappa}}{i\omega\mu} \sum_{m\ne j}^M k^2 g(x_j,x_m)\tau A_m h_m + (\tau A_m \cdot \nabla_x)\nabla g(x,x_m)|_{x=x_j} h_m.
\ee
This linear system can be solved directly using Gaussian elimination method. Then $E$ can be computed as follows
\be \label{eq3.1.2}
    E_{em}=E_{0m}-\frac{ca^{2-\kappa}}{i\omega\mu}\sum_{j\ne m}^M [\nabla g(x,x_j)|_{x=x_m}, \tau A_j]h_j.
\ee
Matrix $\tau$ can be computed as follows
\be \label{eq1.3.4}
        \tau:=I_3-b=\frac{2}{3}I_3, \quad b=(b_{jm}):=\frac{1}{|S|}\int_S N_j(s)N_m(s)ds.
\ee

Let $A_i:=(X_i,Y_i,Z_i)$ then one can rewrite \eqref{eq3.1.1} as
\begin{align}
    &F_x(i)=X_i+\sum_{j\neq i}^M a_{ij}X_j + \sum_{j\neq i}^M b_{ij}Y_j + \sum_{j\neq i}^M c_{ij}Z_j, \label{eq3.1.7}\\
    &F_y(i)=Y_i+\sum_{j\neq i}^M a'_{ij}X_j + \sum_{j\neq i}^M b'_{ij}Y_j + \sum_{j\neq i}^M c'_{ij}Z_j, \label{eq3.1.8}\\
    &F_z(i)=Z_i+\sum_{j\neq i}^M a''_{ij}X_j + \sum_{j\neq i}^M b''_{ij}Y_j + \sum_{j\neq i}^M c''_{ij}Z_j, \label{eq3.1.9}
\end{align}
where by the subscripts $x,y,z$ the corresponding coordinates are denoted, e.g. $F(i)=(F_x,F_y,F_z)(i)$, $F(i):=A_{0i}$, and
\begin{align}
    &a_{ij}:=[k^2 g(i,j) + \partial{x} \nabla g(i,j)_x]D_j,\\
    &b_{ij}:=\partial{y} \nabla g(i,j)_xD_j,\\
    &c_{ij}:=\partial{z} \nabla g(i,j)_xD_j,\\
    &a'_{ij}:=\partial{x} \nabla g(i,j)_yD_j,\\
    &b'_{ij}:=[k^2 g(i,j) + \partial{y} \nabla g(i,j)_y]D_j,\\
    &c'_{ij}:=\partial{z} \nabla g(i,j)_yD_j,\\
    &a''_{ij}:=\partial{x} \nabla g(i,j)_zD_j,\\
    &b''_{ij}:=\partial{y} \nabla g(i,j)_zD_j,\\
    &c''_{ij}:=[k^2 g(i,j) + \partial{z} \nabla g(i,j)_z]D_j,
\end{align}
in which $D_j:=\frac{2}{3}\frac{c a^{2-\kappa}}{i\omega\mu}h_j$.

The error of the method presented in section \ref{sec2} can be estimated as follows. From the solution $E$ of the electromagnetic scattering problem by many small bodies given in \eqref{eq2.1.13}
\be
    E(x)=E_0(x)+\sum_{m=1}^M \nabla\times \int_{S_m} g(x,t) J_m(t)dt,
\ee
we can rewrite it as
\be
    E(x)=E_0(x)+\sum_{m=1}^M [\nabla g(x,x_m), Q_m]+\sum_{m=1}^M \nabla\times \int_{S_m} [g(x,t)-g(x,x_m)] J_m(t)dt.
\ee
Comparing this with the asymptotic formula for $E$ when $a \to 0$ given in \eqref{eq2.1.22}
\be
    E(x)=E_0(x)+\sum_{m=1}^M [\nabla g(x,x_m), Q_m],
\ee
we have the error of this asymptotic formula is
\be
    \text{Error}=\left|\sum_{m=1}^M \nabla\times \int_{S_m} [g(x,t)-g(x,x_m)] J_m(t)dt\right| \sim \frac{1}{4\pi}\left(\frac{ak^2}{d}+\frac{ak}{d^2}+\frac{a}{d^3}\right)\sum_{m=1}^M|Q_m|,
\ee
where $d=\min_m|x-x_m|$ and
\be
    Q_m:=\int_{S_m}J_m(t)dt \simeq -\frac{\zeta_m|S_m|}{i\omega\mu}\tau_m \nabla\times E_{em}=-\frac{\zeta_m|S_m|}{i\omega\mu}\tau A_m, \quad a \to 0,
\ee
because
\be
    |\nabla[g(x,t)-g(x,x_m)]| = O\left(\frac{ak^2}{d}+\frac{ak}{d^2}+\frac{a}{d^3}\right), \quad a=\max_m|t-x_m|.
\ee

To illustrate the idea, consider a domain $\Omega$ as a unit cube placed in the first octant such that the origin is one of its vertex. This domain $\Omega$ contains $M$ small bodies. The small bodies are particles which are distributed uniformly in the unit cube. The following physical parameters are used to solve the problem
\begin{itemize}
     \item Speed of wave, $c=3.0E+10$ cm/sec.
     \item Frequency, $\omega=1.0E+14$ Hz.
     \item Wave number, $k = 2.094395e+04$ cm$^{-1}$.
     \item Constant $\kappa = 0.9$.
     \item Volume of the domain $\Omega$ that contains all the particles, $|\Omega| = 1$ cm$^3$.
     \item Direction of plane wave, $\alpha = (1,  0,  0)$.
     \item Vector $\mathcal{E} = (0,  1,  0)$.
     \item Function $N(x)=Ma^{2-\kappa}/|\Omega|$ where $M$ is the total number of particles and $a$ is the radius of one particle.
     \item Function $h(x) = 1$.
     \item Function $\mu(x) = 1$.
     \item The distance between two neighboring particles, $d = 1/(b-1)$ cm, where $b$ is the  number of particles on a side of the cube.
     \item Vector $A_0$: $A_{0m}:=A_{0}(x_m)=\nabla \times E_0(x)|_{x=x_m}=\nabla \times \mathcal{E} e^{ik\alpha\cdot x}|_{x=x_m}$.
\end{itemize}
The radius $a$ of the particles is chosen variously so that it satisfies and dissatisfies the assumption $ka \ll 1$. The numerical solutions to EM wave scattering problem by many small impedance bodies are computed for $M=125$ and $1000$ particles.

\begin{figure}[htbp]
    \centering
    \includegraphics[scale=0.9]{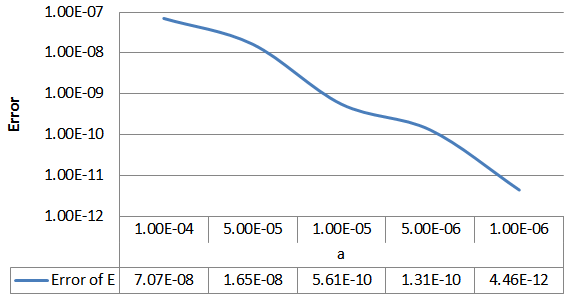}
    \caption{Error of the asymptotic solution $E$ when $M=125$ and $d=2.50E-01$ cm.}
    \label{fig1}
\end{figure}

Table \ref{tab1} and Figure \ref{fig1} show the results of solving the electromagnetic wave scattering  problem with $M=125$ particles, the distance between neighboring particles is $d=2.50E-01$ cm, and with different radius $a$ of particles. When the radius of particles decreases from $1.0E-4$ cm to $1.0E-6$ cm, the error of the asymptotic solution decreases rapidly from about $7.07E-08$ to $4.46E-12$. Note that when $a=1.0E-4$ cm, $ka>1$, which does not satisfy the assumption $ka\ll 1$. When $ka<1$, the error of the solution is always less than $7.0E-08$. This means that the asymptotic formula \eqref{eq2.1.23} for the solution $E$ is perfectly applicable.

\begin{figure}[htbp]
    \centering
    \includegraphics[scale=0.9]{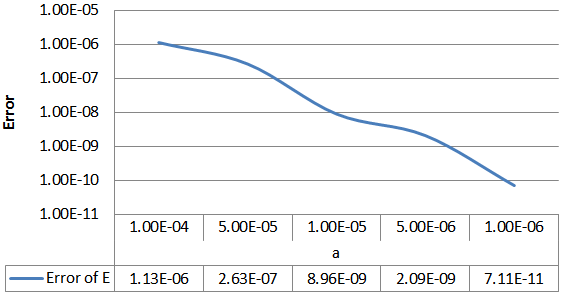}
    \caption{Error of the asymptotic solution $E$ when $M=1000$ and $d=1.25E-01$ cm.}
    \label{fig2}
\end{figure}

\begin{table}[htbp]
  \centering
  \caption{Error of the asymptotic solution $E$ when $M=125$ and $d=2.50E-01$ cm.}
    \begin{tabular}{rrrrrr}
    \toprule
        \multicolumn{6}{c}{M=125, d=2.50E-01} \\
        \midrule
        a     		& 1.00E-04 & 5.00E-05 & 1.00E-05 & 5.00E-06 & 1.00E-06 \\
        Norm of E 	& 1.12E+01 & 1.12E+01 & 1.12E+01 & 1.12E+01 & 1.12E+01 \\
        Error of E 	& 7.07E-08 & 1.65E-08 & 5.61E-10 & 1.31E-10 & 4.46E-12 \\
        \bottomrule
    \end{tabular}%
  \label{tab1}%
\end{table}%
Table \ref{tab2} and Figure \ref{fig2} show the results of solving the problem with $M=1000$ particles, when the distance between neighboring particles is $d=1.25E-01$ cm, and with different radius $a$. From table \ref{tab2}, one can see that the error of the asymptotic solution is also very small when $ka \ll 1$, less than $1.13E-06$, but greater than the previous case when $M=125$ for corresponding $a$. This time, the error is also decreasing rather quickly when the radius of the particles decreases from $1.0E-4$ cm to $1.0E-6$ cm. The small error when $ka \ll 1$ guarantees that the asymptotic formula \eqref{eq2.1.23} for the solution $E$ is well applicable under this assumption.
\begin{table}[htbp]
  \centering
  \caption{Error of the asymptotic solution $E$ when $M=1000$ and $d=1.25E-01$ cm.}
    \begin{tabular}{rrrrrr}
      \toprule
        \multicolumn{6}{c}{M=1000, d=1.25E-01} \\
        \midrule
        a     		& 1.00E-04 & 5.00E-05 & 1.00E-05 & 5.00E-06 & 1.00E-06 \\
        Norm of E 	& 3.16E+01 & 3.16E+01 & 3.16E+01 & 3.16E+01 & 3.16E+01 \\
        Error of E 	& 1.13E-06 & 2.63E-07 & 8.96E-09 & 2.09E-09 & 7.11E-11 \\
      \bottomrule    
    \end{tabular}%
  \label{tab2}%
\end{table}%

\section{Conclusions} \label{sec4}

This paper presents a numerical method for solving EM wave scattering problem by many small impedance bodies. For illustration, the problem is solved with 125 and 1000 particles. The solutions to this problem are computed numerically and asymptotically using the described method. From these experiments, one can observe that if the assumption $ka \ll 1$ is satisfied, the asymptotic formula \eqref{eq2.1.23} yields the solution to the EM wave scattering problem by many small impedance bodies with little error, less than $1\%$. This shows the applicability of the asymptotic method described in the paper.

\bibliographystyle{ieeetran}

\end{document}